\documentclass[trackchanges,twocolumn]{aastex701}
\shorttitle{Jet forecast}
\shortauthors{T.~Kawaguchi \& K.~Yamaoka}

\begin{document}

\title{Jet Forecast: Timing of Discrete Jet Ejections of Stellar-Mass Black Holes Anticipated via X-ray Properties}

\author[orcid=0000-0002-3866-9645,gname=Toshihiro,sname=Kawaguchi]{Toshihiro Kawaguchi}
\affiliation{Graduate School of Science and Engineering, University of Toyama, Gofuku 3190, Toyama 930-8555, Japan}
\email[show]{kawaguti@eng.u-toyama.ac.jp}  

\author[orcid=0000-0003-3841-0980,gname=Kazutaka,sname=Yamaoka]{Kazutaka Yamaoka}
\affiliation{Institute for Space-Earth Environmental Research (ISEE), Nagoya University, Furo-cho, Chikusa-ku, Nagoya, Aichi 464-8601, Japan}
\email{yamaoka@isee.nagoya-u.ac.jp}

\begin{abstract}

The 
radio -- soft $\gamma$-ray 
data 
of the Galactic black hole binary XTE~J1859+226 
showed that 
discrete jets are launched 
when the inner radius of the accretion disk 
($R_{\rm in}$) rapidly approaches the innermost stable 
circular orbit (ISCO). 
Given 
the correlations among the X-ray 
fractional variability 
amplitude (rms), $R_{\rm in}$, and the X-ray hardness, 
previous 
phenomenological findings 
related to jet ejections 
(``jet line'' and sharp drops of rms) 
also indicate that the rapid shrinkage 
of $R_{\rm in}$ down to the ISCO turns on the jet formation.
In order to make clear 
how these 
properties 
appear 
in other objects, 
we investigated the X-ray and radio data of 
XTE~J1550$-$564, GX~339$-$4, and MAXI~J1820+070. 
We found that all the four sources show drops of $R_{\rm in}$ and rms 
1-10\,days 
ahead of a jet activity, 
and exhibit no huge jumps of the accretion rates 
prior to jet ejections. 
XTE~J1859 and MAXI~J1820 show the jet activities 
 under sub-Eddington accretion. 
Models for the jet production that require 
an 
enhancement of $\dot{M}$ or 
a super-Eddington accretion rate are unlikely. 
Another behavior common to multiple sources is the curved shape of the 
rms--$R_{\rm in}$ correlation. 
Therefore, monitoring of $R_{\rm in}$ seems to be 
suitable 
to predict the timing of jet ejections. 
When both $R_{\rm in}$ and rms continuously decrease and when 
$R_{\rm in}$ is about to reach the ISCO, 
alerts can be circulated 
as a ``jet forecast'' for triggering Target-of-Opportunity observations. 


\end{abstract}



\section{Introduction} \label{sec:intro}

Relativistic jet ejections are ubiquitous 
when the 
gas accretes %
onto compact objects, 
especially onto 
black holes (BHs),
from stellar masses \citep{jet2,Belloni2010} 
to supermassive ones 
\citep{Kameno2000,M87jet}. 
The intermediate-mass black hole candidate, 
ESO 243-49 HLX-1 
\citep{Godet2012}, also shows the transient radio emission 
\citep{Webb2012}. 
Single sources occasionally exhibit jet ejections 
in multiple epochs \citep[e.g.,][]{radio,Kuzmicz2017,Dabhade2025}, 
indicating that there is a certain condition/rule for 
launching jets. 

Although more than a century has passed since the discovery of 
jets \citep{Curtis1918}, 
the launching mechanisms are still unclear. 
Rotational energy of the central BHs \citep{Penrose,BZ1977}
and/or the accretion disk \citep{Blandford1982}
can be 
the energy source of jets. 
The configuration of the magnetic field \citep[e.g.,][]{Penna2010}
or the BH masses \citep{Laor2000}
may play a role in the jet formation \citep[for a review, see][]{CzernyYou2016}.  
Supermassive BHs tend to show more relativistic 
jets (with larger Lorentz factors) 
than stellar-mass BH binaries do \citep[e.g.,][]{Vermeulen1994}.

Many attempts have been made 
to observe the moments of the jet ejections, 
and to derive 
constraints on 
plausible theoretical models. 
Detailed observational data 
with dense cadence 
are obtained only for 
some rare cases 
\citep{diskjet_maxij1820, Mirabel1998,Russell2019}, 
e.g., when a source stays long at 
the bright-Hard state 
just before a jet production. 
Since 
it is unforeseeable 
when a jet emerges, 
the 
accumulation of ample 
data around 
the timing of jet ejections is slow. 

\begin{deluxetable*}{lcccccc}
\tablewidth{0pt}
\tablecaption{Properties of the four sources \label{tab:sources}}
\tablehead{
\colhead{Name} & 
\colhead{BH mass} & 
\colhead{Distance} & 
\colhead{Inclination} & 
\colhead{Reference} & 
\colhead{Eddington accretion rate} & 
\colhead{Maximum $T_{\rm in}$}\\
\colhead{} & 
\colhead{($M_{\odot}$)} & 
\colhead{(kpc)} & 
\colhead{(deg)} & 
\colhead{} & 
\colhead{($10^{18}$\,g\,s$^{-1}$)} & 
\colhead{(keV)}
}
\startdata
XTE~J1859+226 & 7.8$\pm$1.9 & 8 & 66.6$\pm$4.3 & 1 & 17.5 & 0.94 \\
XTE~J1550--564 & 9.10$\pm$0.61 & 4.38$^{+0.58}_{-0.41}$ & 74.7$\pm$3.8 & 2 & 20.4 & 0.91 \\
GX~339--4 & 9.8 & 8$\pm$2 & 40 & 3 & 21.9 & 0.90 \\
MAXI~J1820+070 & 9.2$\pm$1.3 & 2.96$\pm$0.33 & 63$\pm$3 & 4 & 20.6 & 0.66 \\
\enddata
\tablecomments{
The inclination angle of 0\,degree corresponds to 
a face-on view. 
References: 
1 \citep{opt_mass}, 
2 \citep{Orosz2011}, 
3 \citep{Connors2019}, 
4 \citep{Atri2020}. 
}
\end{deluxetable*}

There are 
two, conventional arguments on 
the timing of jet ejections. 
First, a jet is launched when a source crosses 
the ``jet line'' by changing the spectral states 
(from the Hard to the Soft states) 
leftward in the Hardness-Intensity diagram 
\citep[HID;][]{jet,Belloni2010}. 
In this work, we focus on the large-scale, transient
jets observed in the state transitions, rather than
compact jets seen only in the Hard-state. 
It is not straightforward to forecast 
a jet production based on the diagram, as follows. 
There is no single, fixed line, 
over which all the jet events pass 
in the HID 
\citep{jet2,Yamaoka2025}. 
Moreover, 
many epochs 
located 
near the ``jet line''
do not pass leftward by the next exposure. 
Therefore, from the position in the HID, 
the timing of the coming jet is unpredictable.

Next, 
the X-ray fractional 
variability amplitude (rms), 
as well as the emergence of the Type-B QPOs \citep{jet2}, 
is also related to the jet timing. 
For several objects, sharp drops of rms 
are associated with radio activity within a few days \citep{jet2}. 
However, 
it is still unclear 
to what extent the variability amplitude must be reduced for launching jets.

\cite{Yamaoka2025} 
then searched for conditions (or precursors) of such discrete jet events, 
by focusing on an X-ray binary system, 
XTE~J1859$+$226 \citep{discovery,rxte99}. 
This object 
showed 
several discrete jet ejections on a 
short timescale (
$\sim 30$\,days) around 
the peak of the X-ray brightness \citep{radio, Yamaoka2025}, 
and thus is suitable for investigating 
which quantity shows a characteristic behavior before jet ejections.
Jets 
turn out to emerge 
when the following 
two conditions are satisfied. 
First, the inner radius of the accretion disk 
($R_{\rm in}$) 
shrinks sufficiently fast. 
Secondary, it reaches the innermost stable 
circular orbit (ISCO). 

They also found 
the correlations among 
rms, 
$R_{\rm in}$, and the X-ray hardness. %
As 
$R_{\rm in}$ 
gets smaller, 
less X-ray variability and softer X-ray spectra are realized. 
The X-ray variability is controlled by the competition between the relatively stable disk and the highly-variable hard component. 
Therefore, 
the two 
characteristic behaviors around 
jet ejections of 
a number of BH 
sources 
(``jet line'' and sharp drops of rms) 
seem to be physically 
caused by the time variation of $R_{\rm in}$.

Although 
the conditions for jet launching 
(i.e., the shrinkage of $R_{\rm in}$) 
and 
the tight correlation between rms and $R_{\rm in}$ 
are discovered 
\citep{Yamaoka2025}, 
it is investigated 
only for one source so far. 
In order to make clear 
if the same or similar conditions (or precursors) 
operate 
commonly, 
we here investigate 
XTE~J1550$-$564, GX~339$-$4, and MAXI~J1820$+$070 that 
have relatively rich X-ray and radio data around the jet ejections. 
All the four sources have small column densities, 
with an order of $10^{21}$cm$^{-2}$ \citep{Yamaoka2025,Steiner2011,Shidatsu2018,WangJi2018}.

For the four sources, 
the BH masses, 
distances, inclination angles of the binary orbits, 
 adopted in this work, 
and 
relevant 
references 
are listed 
in Table~\ref{tab:sources}. 
Rightmost column shows the maximum temperature 
of the disk ($T_{\rm in}$) obtained via the 
X-ray fitting (\S~\ref{sec:data}). 
For a nonrotating BH, the ISCO radius 
is 3 times the Schwarzschild radius $R_{\rm Sch}$ 
[$3 \, R_{\rm Sch} \approx 88.6 \, (M_{\rm BH} / 10 M_{\odot} )$\,km]. 
The Eddington accretion rate ($ 16 \, L_{\rm Edd} /c^2 $, assuming 
the efficiency from 
the rest mass energy of the accreting gas to radiation being 1/16) 
is $2.24 \times 10^{19} ( M_{\rm BH} / 10 M_{\odot} )$\,g\,s$^{-1}$, 
where 
the Eddington luminosity $L_{\rm Edd}$ is 
$1.26 \times 10^{39} ( M_{\rm BH} / 10 M_{\odot} )$\,erg\,s$^{-1}$.

In this work 
we report the time variation of the 
X-ray spectral and temporal properties and the 
accretion rate for the four sources. 
The statistical errors are quoted for 68\,\% confidence level. 
In the next section we briefly describe the data we use and 
the analysis methods for the X-ray data. 
Then, results for individual sources are briefly summarized. 
In section~\ref{sec:common} we present the behaviors 
that are commonly seen in multiple sources. 
Finally, we
summarize this study with some discussions 
in section~\ref{sec:summary}.

\section{Data and analysis methods} \label{sec:data}

The X-ray data of XTE~J1859 at 2--250\,keV, obtained via RXTE/PCA and HEXTE, 
and the radio data 
were analyzed 
by \cite{Yamaoka2025}. 
We use their fitting results in this work. 
We also analyze the RXTE/PCA+HEXTE data 
for XTE~J1550 in the 1998 outburst \citep{Sobczak1999}
and GX~339 in the 2002-2003 outburst \citep{Belloni2005}, 
with HEASoft v6.36. 
For MAXI~J1820, we use 
the 
MAXI/GSC data 
\citep{Shidatsu2019} 
and the NICER data 
at 2--20\,keV and 0.3--10\,keV energy bands, respectively, 
and we fit the two datasets separately in order to see 
the consistency (\S~\ref{sec:indiv}). 
The radio light curves 
at 843\,MHz and 4.8\,GHz 
for XTE~J1550 and GX~339
were 
reported by \cite{Wu2002} and \cite{Gallo2004}, respectively, 
while detailed radio data 
are presented for MAXI~J1820 by 
\cite{Bright2020}.

We analyze the X-ray data in the 
standard 
manner as done by \cite{Yamaoka2025}, 
for the four 
 sources in total 
(XTE~J1859, XTE~J1550, GX~339, and MAXI~J1820). 
Below, we briefly summarize 
the spectral and timing properties that we use.

There are the following two components. 
From a multi-temperature accretion disk 
[{\tt diskbb} \citep{mcd1,mcd2} in XSPEC,  
convolved with the SIMPL Comptonization \citep{simpl}], 
we deduce the temperature of the inner disk ($T_{\rm in}$), 
the disk luminosity ($L_{\rm disk}$), 
and the inner radius ($R_{\rm in}$). 
Considering the inner boundary 
of the disk 
and the spectral hardening due to electron scattering, 
a better estimation of  
$R_{\rm in}$ may be 1.19 times the $R_{\rm in}$ derived 
above \citep{bhmass_est}. 
Throughout this work, 
we mention the 
$R_{\rm in}$ derived via the X-ray fitting 
without the correction 
for simplicity. 

A Comptonized, power-law component ({\tt simplcutx}) 
consumes a part of the disk photons. 
This component represents a hot 
plasma with an electron temperature of $\sim 100$\,keV, 
such as the corona above and 
below the disk or an advection-dominated accretion flow [ADAF; 
\cite{ADAF,ADAF2}].
For MAXI~J1820, there are no clear 
narrow Fe~K$\alpha$ line nor 
absorption features in the NICER data, and hence 
they are not added in the model fitting. 

Another critical quantity 
is the X-ray 
fractional variability amplitude (rms) 
calculated over the frequency range of 0.03--64\,Hz of 
the power spectral densities \citep{qpodef2}.

\section{Results for individual sources} \label{sec:indiv}

\begin{figure}
\epsscale{1.1}
\plotone{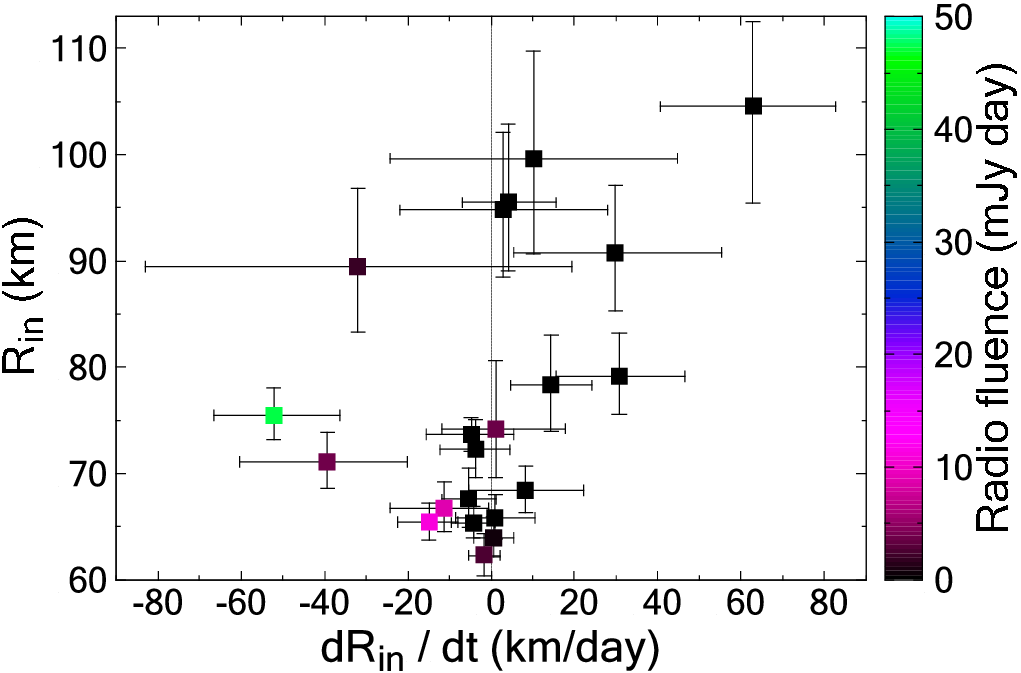}
 \caption{
 $R_{\rm in}$ (at the latter epoch of the two epochs for calculating $dR_{\rm in}/dt$) 
as a function of $dR_{\rm in}/dt$. 
The color bar shows the one-day radio 
fluence 
for each data point \citep{Yamaoka2025}. 
 }
\label{fig:rin_drdt_radio_xtej1859}
\end{figure}

{\it XTE~J1859}: 
At the beginning of the observations, 
the disk inner radius 
was far away from the BH \citep{Yamaoka2025}. 
Then, it approaches the BH (down to $\sim$\,64\,km), 
increasing 
both 
$T_{\rm in}$ and $L_{\rm disk}$.
They fit the radio light curve by a form 
comprising five 
Fast-Rise-Exponential-Decay flares. 
Figure~\ref{fig:rin_drdt_radio_xtej1859} shows 
$R_{\rm in}$ as a function of $dR_{\rm in}/dt$, 
with 
different colors meaning 
the one-day radio 
fluence of a coming radio flare. 
The time variation of 
$R_{\rm in}$, 
obtained by the current data, 
is an order of $| dR_{\rm in}/dt | 
\lesssim 50$~km/day. 
It is much slower than the radial velocity of 
the standard accretion disk 
at 10\,$R_{\rm Sch}$ ($\sim 2 \times 10^6$\,km/day; 
for 
a half of the 
Eddington accretion rate 
and the viscosity parameter $\alpha$ of 0.03) and 
than the Keplerian rotational velocity 
at the same radius 
($\sim 6 \times 10^9$\,km/day). 
Data points with the large radio fluence are 
mostly 
concentrated in 
the lower-left corner, 
meaning that jets are ejected when $R_{\rm in}$ rapidly shrinks 
(to $\sim$\,64\,km)
and is about to reach the ISCO. 
Little (or no) radio activity is realized when $dR_{\rm in}/dt$ is zero or positive 
or when $R_{\rm in}$ is already at the ISCO.

The accretion rate $\dot{M}$ 
can be estimated 
via the $L_{\rm disk}$--$R_{\rm in}$ diagram \citep{Yamaoka2025}, 
given that 
$L_{\rm disk}$ 
is likely in proportion to $\dot{M} / R_{\rm in}$ 
[$L_{\rm disk} = (1/2) (G M_{\rm BH} \dot{M} /R_{\rm in})$; \cite{standard}]. 
If the hard component is emitted from a vertically 
located 
region 
[i.e., corona in the Soft state; \cite{ADAF2}],   
the accretion rate at that region is also relevant in principle. 
Unlike active galactic nuclei, the hard component 
in the Soft-state of stellar-mass BHs is noticeably 
fainter than the disk. 
Thus, we consider only $L_{\rm disk}$ for 
the $\dot{M}$-estimation. 
When the hard component and the disk are located 
horizontally, such as the ADAF inside 
$R_{\rm in}$ of the truncated disk during the Hard-state, 
$\dot{M}$ estimated via $L_{\rm disk}$ likely represents 
the accretion rate of the system. 
\cite{Yamaoka2025}
showed that there is no sudden 
nor large enhancement of $\dot{M}$ at the 
timing of jet ejections of XTE~J1859.

\begin{figure*}
\epsscale{1.1}
\figurenum{2}
\plotone{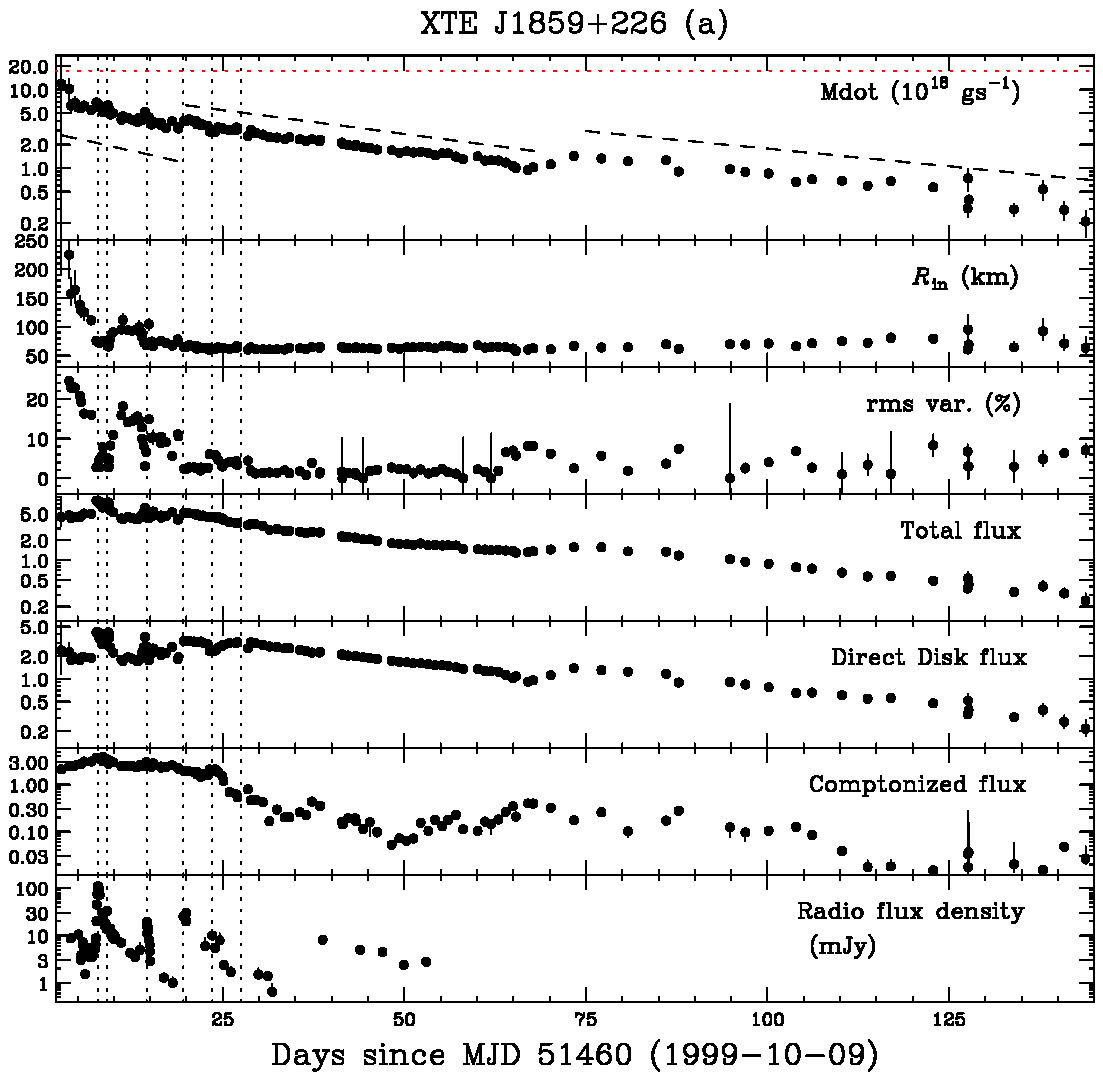}
 \caption{
 Time evolution of $\dot{M}$ and other quantities 
 for the whole observing epochs over $\sim$140~days 
(a) and over the epochs relevant to the jet ejections (b) for 
XTE~J1859. 
Units for the total, direct disk, and Comptonized fluxes are 
$10^{-8}$\,erg\,cm$^{-2}$\,s$^{-1}$.
Horizontal red 
dotted lines are the Eddington accretion rate 
($17.5 \times 10^{18}$\,g\,s$^{-1}$; \S~\ref{sec:intro}). 
Vertical 
dotted 
lines indicate the timing of jet ejections 
[the radio peak times; \cite{Yamaoka2025,radio}], 
which all occur with 
sub-Eddington accretion rates.
Dashed lines show 
the scaling 
of $\dot{M}$ 
with the $e$-folding timescales of 
18, 38, and 46\,days 
fitted for Day intervals at  
4--19, 20--65, and 72--145, 
respectively. 
}
\label{fig:Mdotxtej1859}
\end{figure*}

\begin{figure*}
\epsscale{1.1}
\figurenum{2}
\plotone{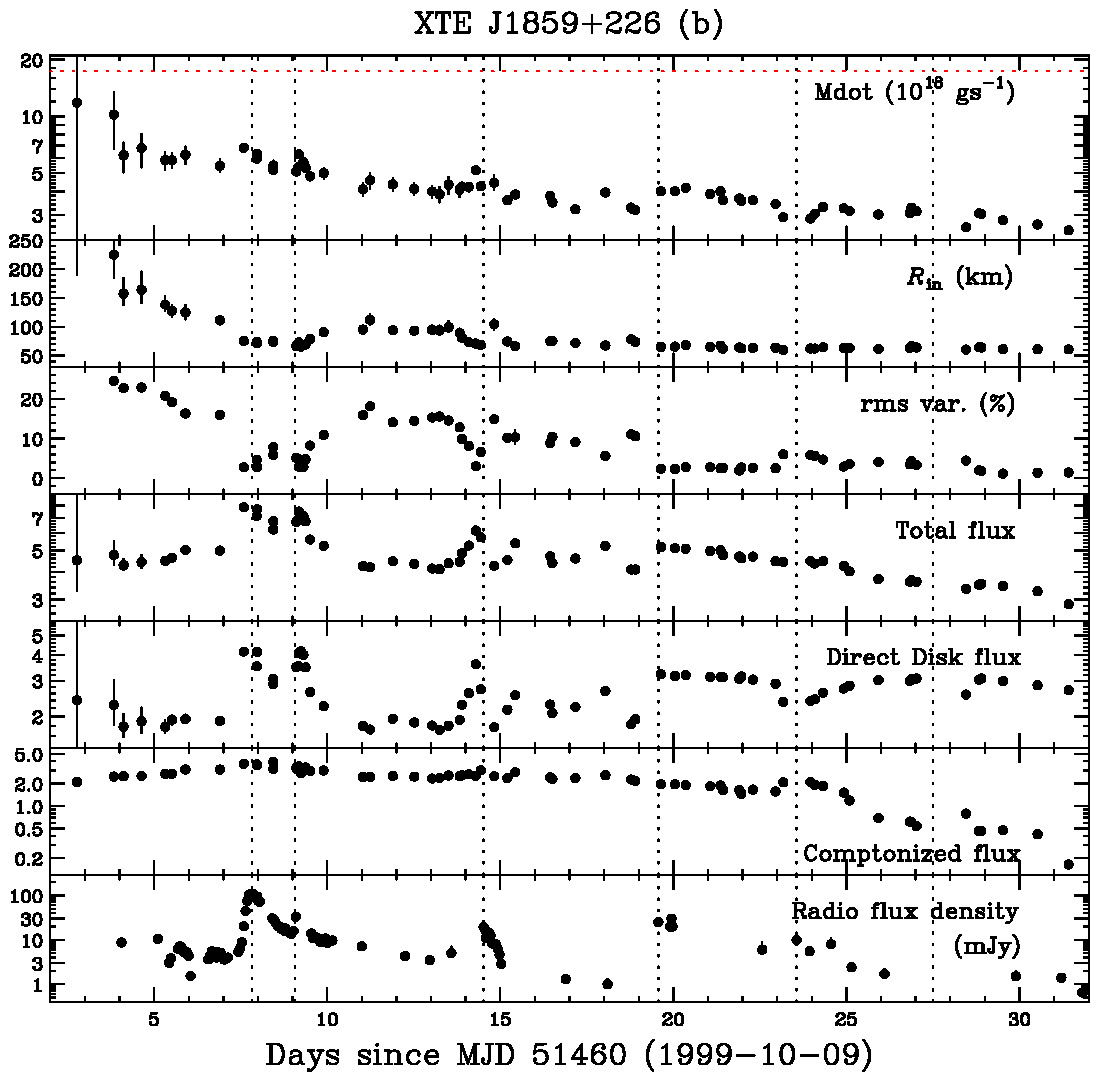}
 \caption{\it Continued
}
\end{figure*}


In this work, we further investigate the time evolution of 
$\dot{M}$ 
[$= 2 \, L_{\rm disk} \, R_{\rm in} / (G M_{\rm BH}) 
= 1.51 \times 10^{18} ( L_{\rm disk} / 10^{38} {\rm erg}\,{\rm s}^{-1} ) 
( R_{\rm in} / 100 {\rm km} ) ( M_{\rm BH} / 10 M_{\odot} )^{-1}$\,g\,s$^{-1}$] 
(figure~\ref{fig:Mdotxtej1859}). 
A correction for $R_{\rm in}$ due to 
the inner boundary and electron scattering 
(\S~\ref{sec:intro}) would result in 
a factor of 1.19 larger value for the estimated $\dot{M}$. 
Even though 
we do not apply the correction in this work, 
qualitative conclusions hereafter are not altered.

Overall, the time sequence of $\dot{M}$ 
shows a gradual, exponential decay 
after the 
start of observations, 
with an $e$-folding timescale of $29.6 \pm 1.5$\,day over 
the whole $\sim$140\,day data. 
The $e$-folding timescale seems to increase progressively with time. 
There are slight shifts in the normalization of 
the exponential decay 
at Day$\sim$20 and $\sim$70, resembling the 
secondary outbursts (reflares) seen in BH sources \citep{Tanaka1992}.

We see no huge jumps of $\dot{M}$ ahead of jet 
ejections. 
Although there is a hint 
of $\sim$20\% enhancement of $\dot{M}$ just before 
the jet productions (at Day $\sim$7.5, 9, 14.5, 19.5), 
the current data are insufficient 
for 
reliable evaluation. 
While 
the X-ray flux 
increases 
very quickly in the first $\sim$10\,days 
(with the peak luminosity of $0.6L_{\rm Edd}$ at Day~8.0), 
$\dot{M}$ 
decreases 
in the 
brightening phase \citep{Yamaoka2025}. 
The ratio of the estimated $\dot{M}$ over the 
Eddington accretion rate (\S~\ref{sec:intro}) is 
proportional to $M_{\rm BH}^{-2}$. 
Jet activities of this source seem to occur 
with sub-Eddington accretion rates, 
unless its true $M_{\rm BH}$ is much smaller by a considerable 
 (e.g., $\gtrsim$\,2) factor.

{\it XTE~J1550}: 
Figure~\ref{fig:rin_rms_flux_xtej1550_gx339_maxij1820} 
shows 
the time evolution of 
$\dot{M}$, 
$R_{\rm in}$, 
rms, 
various fluxes from X-ray data, 
and the radio flux density 
for XTE~J1550, GX~339, and MAXI~J1820.
In the 80-day time series
for XTE~J1550 [panel (a)], 
there is a radio flare 
with the peak at Day$\sim$18 (bottom row). 
The radio peak epoch can, in principle, be 
affected by the synchrotron self-absorption. 
The time-lag between 
the starting time of a radio flare 
and 
the radio peak time 
is 
$\sim$0.1-0.3\,day 
in 
XTE~J1859 and MAXI~J1820 
\citep{Yamaoka2025,diskjet_maxij1820}.

\begin{figure*}
\epsscale{1.1}
\figurenum{3}
\plotone{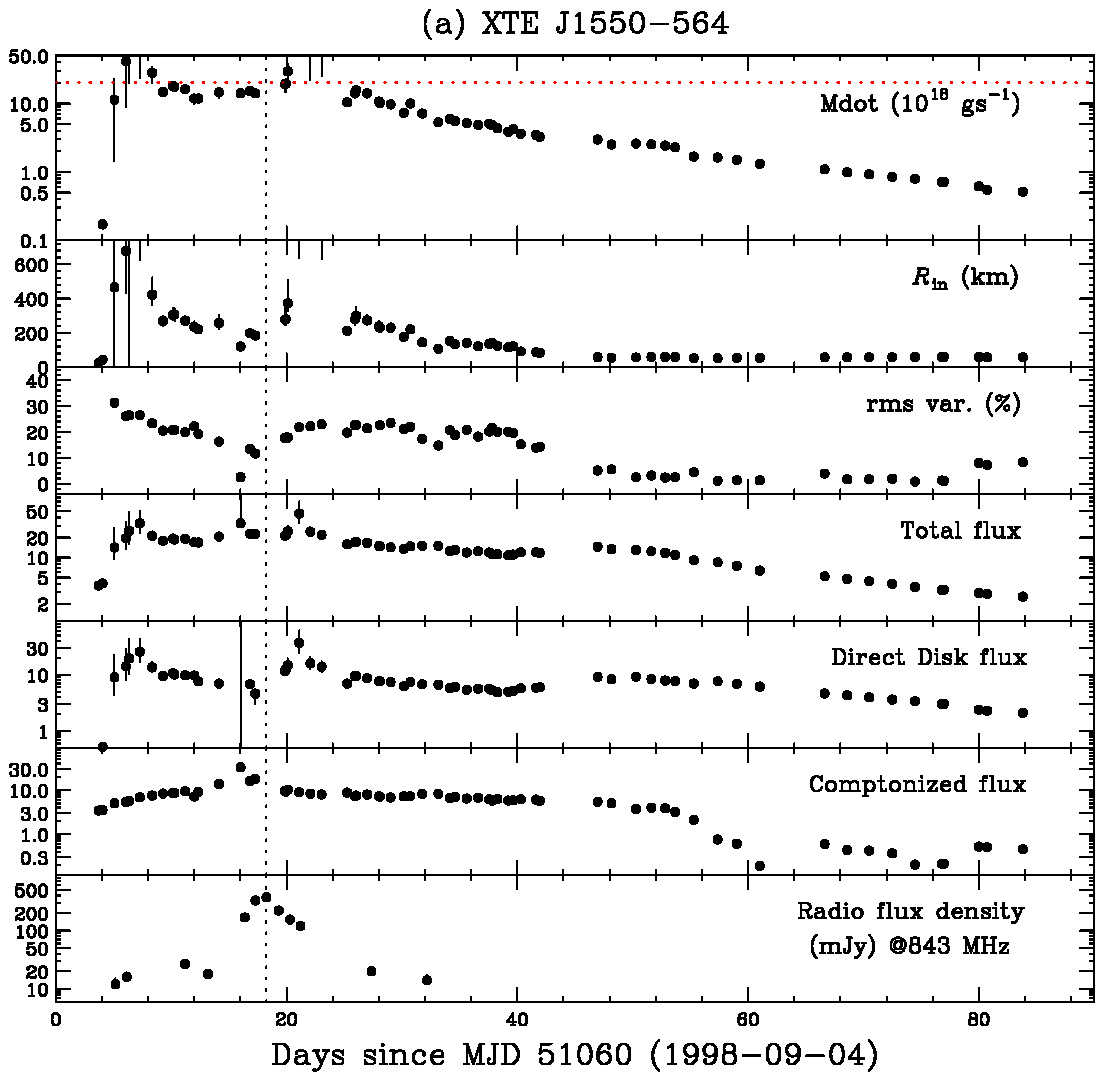}
 \caption{
 Time variation of various quantities over $\sim$80~days 
for XTE~J1550 (a), 
over $\sim$
370~days for GX~339 (b), and 
over {$\sim$}
16~days 
 for MAXI~J1820 (c). 
A couple of radio data points at negative Days are 
also shown for GX~339 \citep{Corbel2013}. 
At the begining of X-ray observations of GX~339, 
$R_{\rm in}$ is not well constrained (i.e., showing 
long eroor-bars) with large fitted values. 
For the X-ray properties of MAXI~J1820, 
black circles 
and red stars 
are obtained via MAXI and NICER, respectively. 
Units for the fluxes are the same as 
those in figure~\ref{fig:Mdotxtej1859}. 
Horizontal red dotted lines 
are 
the Eddington 
accretion rates 
(20.4, 21.9, and 20.6 in the unit of $10^{18}$\,g\,s$^{-1}$ 
for XTE~J1550, 
GX~339, and MAXI~J1820, respectively).
Vertical dotted lines show the timing 
with 
the radio peak flux densities.
}
\label{fig:rin_rms_flux_xtej1550_gx339_maxij1820}
\end{figure*}

\begin{figure*}
\epsscale{1.1}
\figurenum{3}
\plotone{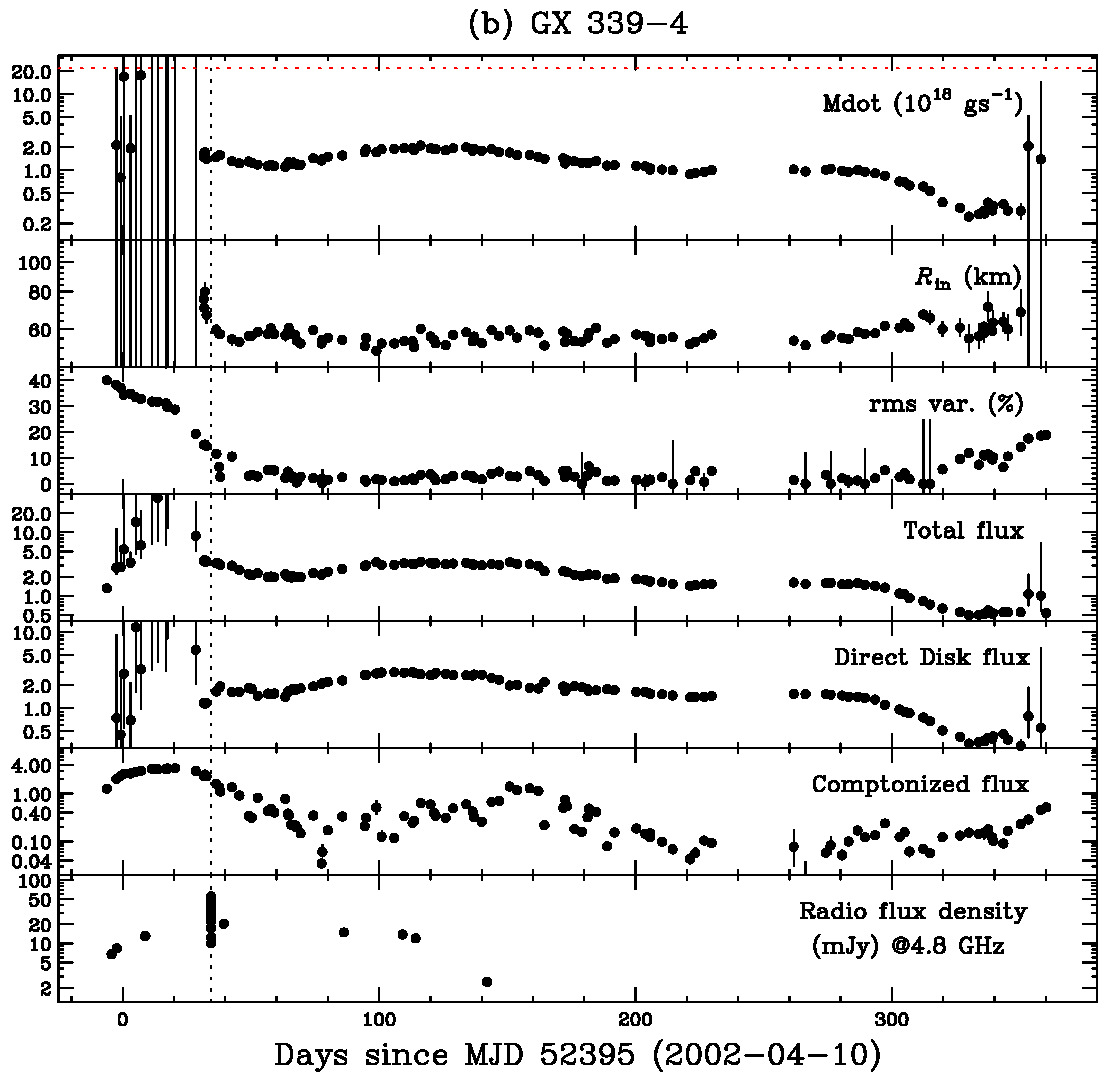}
 \caption{\it Continued
}
\end{figure*}

\begin{figure*}
\epsscale{1.1}
\figurenum{3}
\plotone{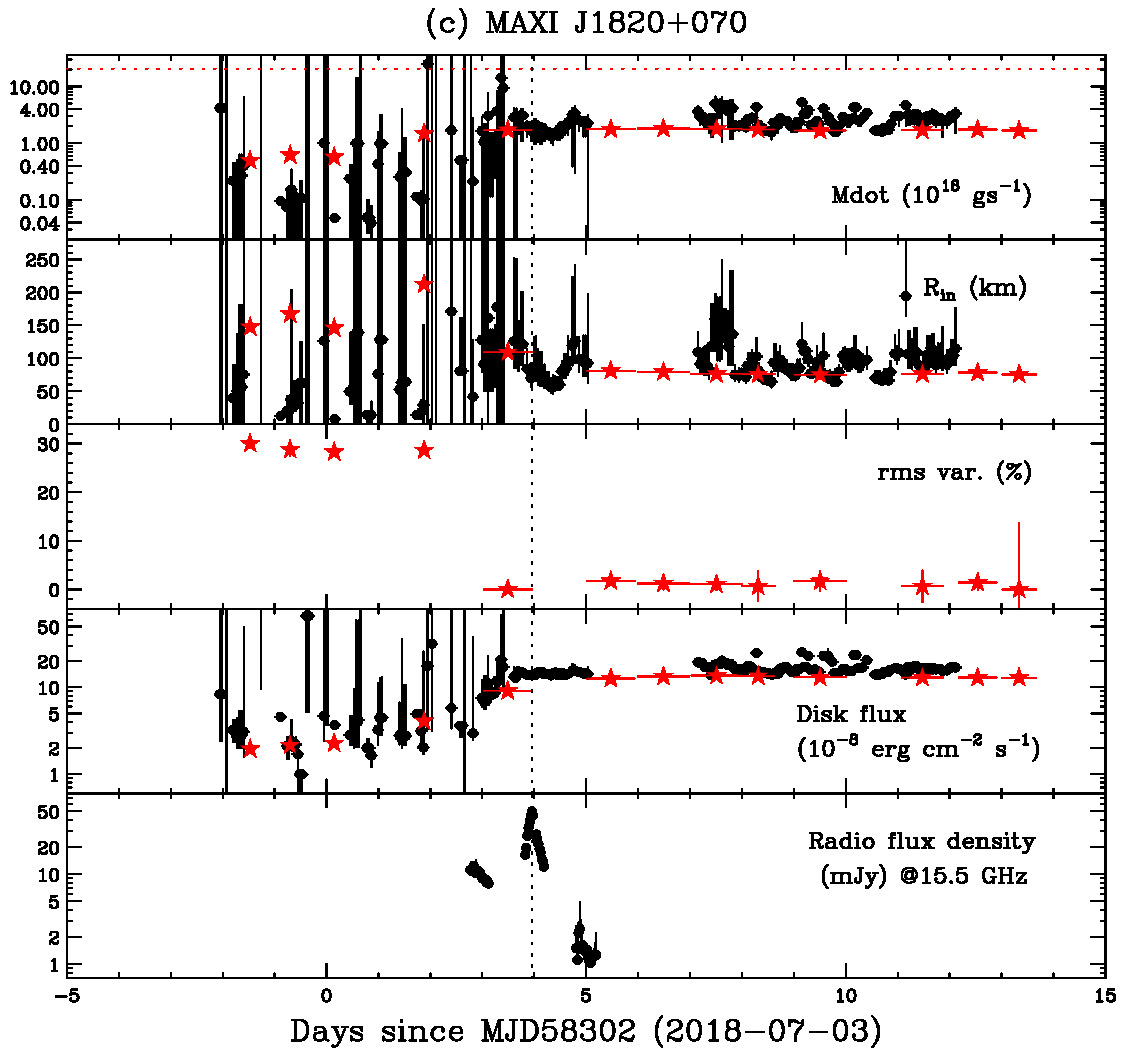}
 \caption{\it Continued
}
\end{figure*}

There is a drop of rms 
(third row) from 
$\sim$5\,days ahead of the radio peak. 
$R_{\rm in}$ starts a rapid shrinkage 
from 
$\sim$10\,days before 
the radio peak epoch, 
and then 
arrives 
at the minimum level 
($\sim$58\,km) 
just before 
the radio peak (second row). 
The ISCO radius 
around a 9.1\,$M_{\odot}$, nonrotating BH is 81\,km. 
Even if a factor of 1.19 correction in $R_{\rm in}$ 
is taken into account, 
there remains a difference to some extent. 
It implies 
either a 
rotating BH \citep[e.g.,][]{Steiner2011} or 
an overestimation of the BH mass 
(Table~\ref{tab:sources}).

Prior to the radio peak, 
the estimated $\dot{M}$ 
shows no 
clear 
rise, 
and stays around the Eddington rate. 
If its $M_{\rm BH}$ is overestimated (as suggested 
above), 
the true ratio of $\dot{M}$ 
over the 
Eddington 
rate 
is enlarged (i.e., super-Eddington accretion).
After the radio activity, $\dot{M}$ 
decays exponentially.
The $e$-folding 
timescale is 10.8\,day during Day~25--35, 
and then 
is 19.8\,day afterwards.

\begin{figure}
\epsscale{1.1}
\figurenum{4}
%
\plotone{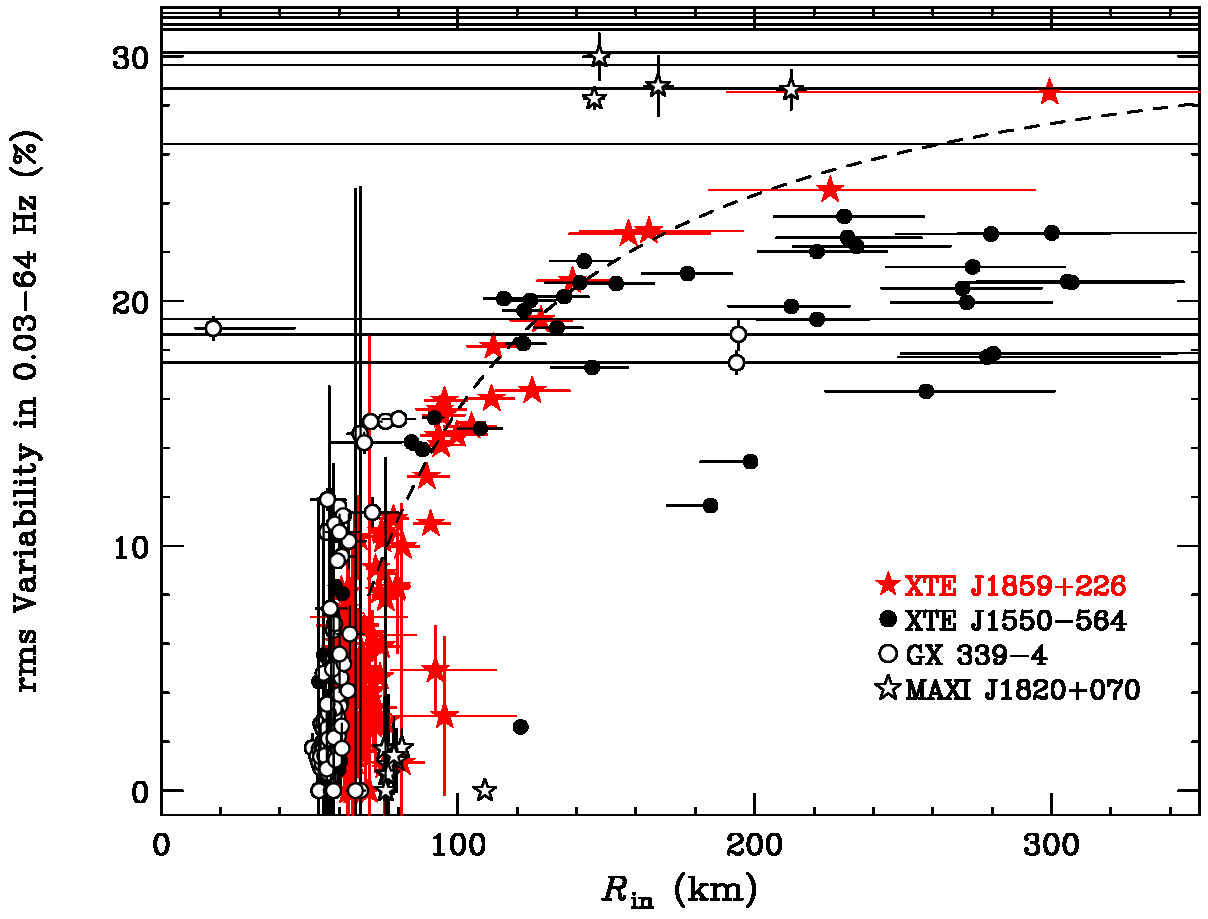}
 \caption{
Relation between $R_{\rm in}$ and rms for XTE~J1550 
 (filled circles), GX~339 
 (open circles), and MAXI~J1820 (open stars).
 Dotted curve is the fitting result 
 for XTE~J1859 [red filled stars; \cite{Yamaoka2025}]. 
 }
\label{fig:rin_rms_all}
\end{figure}

In figure~\ref{fig:rin_rms_all}, 
XTE~J1550 
shows a curved relation between 
rms 
and 
$R_{\rm in}$. 
This source 
is another example (other than XTE~J1859) that shows both a precursor of 
a discrete jet ejection and the positive correlation between 
$R_{\rm in}$ and rms. 
General trends seen in an earlier version of the analyses 
are presented by \cite{Kawaguchi2026}.

{\it GX~339}: 
The radio peak time of GX~339 
(at Day$\sim$35) is covered by a series of 
high-cadence observations \citep{Gallo2004}, which 
are shown as vertically spread data points in 
panel (b) of 
figure~\ref{fig:rin_rms_flux_xtej1550_gx339_maxij1820}. 
The sharp drops of both rms and $R_{\rm in}$ start 
$\sim$10\,days 
before the peak of the radio flux 
density, 
%
and then $R_{\rm in}$ reaches the minimum level 
($\sim$57\,km) 
around the radio peak time. 
It is smaller than 
the ISCO radius 
around a 9.8\,$M_{\odot}$, nonrotating BH (87\,km).

There is no clear enhancement of 
$\dot{M}$ 
ahead of the radio peak in GX~339. 
Although 
the estimated $\dot{M}$ is 
sub-Eddington 
($\sim 1/10$ of 
the Eddington rate), 
the exceptionally large uncertainty of 
$M_{\rm BH}$, distance, and the orbit inclination 
for this source \citep{Connors2019}
hinders a firm conclusion on whether 
it is sub- or super-Eddington accretion. 

The rms--$R_{\rm in}$ relation (figure~\ref{fig:rin_rms_all}) 
exhibits 
a vertical distribution of rms, from 0 to $\sim$15\%, 
staying at similar $R_{\rm in}$. 
When $R_{\rm in}$ is large, away from the ISCO radius, 
rms is also large.

{\it MAXI~J1820}: 
Panel (c) 
of 
figure~\ref{fig:rin_rms_flux_xtej1550_gx339_maxij1820} 
shows 
the results 
for MAXI~J1820.
Here, 
the disk total flux 
(the direct flux plus photons consumed by the 
hard component)
is shown. 
Although the NICER datasets tend to show 
slightly 
smaller $R_{\rm in}$ and 
the disk flux than the MAXI data, 
two datasets roughly show consistent values and time variations.  
During the latter half of the 
observing time, $R_{\rm in}$ stays at a small radius of 
$\sim 76$\,km, 
close to 
the ISCO radius around a 9.2\,$M_{\odot}$, nonrotating BH ($=$\,81\,km). 
About 2\,days before the 
radio peak, 
$R_{\rm in}$ and rms start 
to drop to the minimum level. 
Since MAXI~J1820 has a relatively low-temperature disk 
($T_{\rm in} \leq 0.66$\,keV), 
compared to e.g., 
$\leq 0.94$\,keV 
of XTE~J1859, 
its $T_{\rm in}$ and $R_{\rm in}$ 
(especially during the epochs with large $R_{\rm in}$) 
are difficult to be determined 
precisely 
via the MAXI/GSC data at 2--20\,keV.

The estimated $\dot{M}$ is sub-Eddington 
($\lesssim 1/10$ of the Eddington rate), and shows 
no enhancement prior to the jet ejection. 
Even if 
$M_{\rm BH}$ is overestimated 
by a factor of $\lesssim$3, 
its jet is apparently launched 
under a sub-Eddington accretion rate. 

The rms-$R_{\rm in}$ relation for this source 
(figure~\ref{fig:rin_rms_all}), shown 
for the NICER data, indicates 
a positive correlation, although the number of data points is 
rather small.

\section{Behaviors common to multiple sources} \label{sec:common}

All the four sources turned out to show 
the precursors of jet ejections
1--10 days 
before radio activities. 
Namely, the drops of rms and $R_{\rm in}$ start 
1--2~days ahead of each jet ejection 
for XTE~J1859 \citep{Yamaoka2025}, 
5--10~days and 
10~days 
before the maximum radio flux 
densities 
for XTE~J1550 and GX~339, 
respectively, 
and 
2~days 
ahead of 
a jet ejection for 
MAXI~J1820. 
Better-quality data, which will be obtained via the 
``jet forecast'', would uncover what determines these timescales. 

All the four sources 
show 
no huge jumps of $\dot{M}$ ahead of jet ejections. 
Something else, other than the 
accretion rate, must control the jet production. 
Frameworks for the jet formation 
that require (or always accompany) 
a temporary 
slide of a large mass towards the central BH 
(with a small fraction of the infalling gas being 
turned at a right angle 
and 
accelerated to a relativistic velocity) 
are disfavored by the lack of a large enhancement of $\dot{M}$.
Gas infalling with few increases in $L_{\rm disk}$ 
(and thus in estimated $\dot{M}$), via e.g., subtracting 
the angular momentum of the disk gas by the gas ejections, 
is not contradicted. 
If this is the case, however, 
the depletion of the gas at the vicinity of the 
BH would reduce appreciably 
$\dot{M}$ for a while after a jet ejection, 
which is not observed.

Jet ejections of XTE~J1859 and MAXI~J1820 occur with 
sub-Eddington accretion rates, while it is 
uncertain for XTE~J1550 and GX~339. 
Models for the jet ejection 
that 
need 
super-Eddington accretion rates 
are improbable.

Another behavior common to multiple sources 
(XTE~J1859 and XTE~J1550) 
is the curved (concave upward) shape of the 
rms--$R_{\rm in}$ correlation, 
meaning the following two aspects. 
(1) The shrinkage of $R_{\rm in}$ does 
not accompany 
a sharp drop of rms in the beginning. 
In other words, 
the rapid decrease of $R_{\rm in}$ starts earlier than 
that of rms. 
Later, the $R_{\rm in}$-shrinkage and the drop of rms 
keep step to a jet production. 
In terms of the ``jet forecast'', 
the monitoring of $R_{\rm in}$ is thus a more effective 
tool to notice (and start a careful watch on) 
a coming jet ejection, than 
that of rms. 
(2) The convex-upward shape also means that various rms values 
appear 
when $R_{\rm in}$ stays around the minimum radius. 
It can be a reason why it is still unclear to what 
extent rms must be low for a jet ejection.

For XTE~J1859, 
\cite{Yamaoka2025} 
found a curved relation as  
rms[\%] = $33.0-25.3 (R_{\rm in}/70 [{\rm km}])^{-1.01}$.
For XTE~J1550 
(figure~\ref{fig:rin_rms_all}), the convex is even 
stronger. 
The 
characteristic (or universal) shape 
in the rms--$R_{\rm in}$ relation seen in 
multiple sources indicates that $R_{\rm in}$ may be roughly estimated 
from the X-ray variability amplitude. 
When an epoch falls appreciably off this curve, 
something unusual is undertaking, such as a time-stable 
hot flow or a highly-variable disk.

When both rms and $R_{\rm in}$ continuously decrease and when 
$R_{\rm in}$ is about to reach the ISCO, 
alerts can be circulated 
as a ``jet forecast'' for triggering Target-of-Opportunity (ToO)
observations. 
The timing of the coming 
jet production will be estimated from 
$d R_{\rm in}/dt$ and the known (or inferred) 
ISCO radius.

\section{Summary and discussions} \label{sec:summary}

To investigate what drives the production 
of discrete
jets from the vicinity of BHs, 
we have searched for conditions (or precursors) of 
jet events from BH binaries. 

   \cite{Yamaoka2025}
   took a close look at the X-ray, soft $\gamma$-ray, and 
   radio data obtained for 
   XTE~J1859$+$226. 
They found 
that a rapid shrinkage of the disk inner radius $R_{\rm in}$ 
down to the ISCO is the condition for jet production. 
They 
also found strong correlations among the X-ray variability, 
$R_{\rm in}$, 
and the X-ray hardness. 
Therefore, two phenomenological arguments so far 
on jet productions, ``jet line'' and drops of 
the X-ray variability amplitude rms,  
also indicate that the rapid shrinkage 
of $R_{\rm in}$ down to the ISCO turns on the jet formation.

We examined the time evolution of 
the accretion rate $\dot{M}$ of XTE~J1859.
It turned out to be  
exponentially decaying during the outburst (with an $e$-folding timescale of 
$\sim$30\,days), 
and shows no huge jumps around the 
jet ejections. 
There is a hint of $\sim$20\% enhancement of 
$\dot{M}$ just before 
the jet productions. 
Such fine structures in the time evolution of $\dot{M}$ 
will be examined 
by the better data via 
the ``jet forecast'' in the future.

Next, 
in order to make clear 
if the 
precursors 
of jet ejections 
emerge 
in other objects (other than XTE~J1859), 
we investigated the X-ray and radio data of 
XTE~J1550$-$564, GX~339$-$4, and MAXI~J1820+070. 
We found that multiple sources do 
show the similar behaviors 
in the timing and spectral properties, as follows.

All the four sources turned out to show 
the precursors of jet ejections (drops of $R_{\rm in}$ and rms)
1--10 days 
before radio activities. 
They show no huge jumps of $\dot{M}$ prior to 
each jet ejection. 
Jets in XTE~J1859 and MAXI~J1820 occur under sub-Eddington accretion. 
Theoretical models for the jet production that require 
(or always accompany) 
an upsurge 
of $\dot{M}$ or 
a super-Eddington accretion rate are improbable.

Another behavior common to multiple sources is the characteristic, 
curved shape of the 
rms--$R_{\rm in}$ correlation. 
Both XTE~J1859 \citep{Yamaoka2025} 
and XTE~J1550 
show a convex-upward form in the diagram. 
(1) When $R_{\rm in}$ starts to rapidly decrease, a drop of the rms 
is not yet significant. 
(2) After $R_{\rm in}$ reaches and stays at the ISCO, rms shows 
a variety of amplitudes. 
Namely, (1) it is easier to notice a drop of $R_{\rm in}$ than that of rms 
in the beginning of the precursor. 
(2) It is unclear how much rms must be reduced for launching 
a jet. 
Therefore, monitoring of $R_{\rm in}$ seems to be more suitable 
to predict the timing of jet ejections.

Monitoring of the inner radius, by a sort of pipe-line products, 
to forecast 
the timing of jet production, 
will raise the success rate of 
ToO observations. 
Namely, the (approved) observing time can be used selectively 
in the expected timing of jet productions. 
Then, better quality-data with more frequent cadence 
will be obtained in the future, 
which will unveil, 
e.g., i) how fast 
$d R_{\rm in} / dt$ must be for 
launching a discrete jet, 
and ii) the physics behind how early the precursors start 
prior to each jet ejection.

When both $R_{\rm in}$ and rms continuously decrease and when 
$R_{\rm in}$ is about to reach the ISCO, 
alerts can be circulated 
as a ``jet forecast'' for triggering ToO 
observations. 
The timing of the coming jet production can be estimated from the 
time derivative of $R_{\rm in}$ and the known (or inferred) 
ISCO radius. 
The ISCO radius can be estimated 
from the past time series of 
the radius, 
or from the inner radius at 
a preceding jet ejection. 
    When $R_{\rm in}$ 
    is at the ISCO, 
    jet ejections are not expected for a while, 
    until it leaves the ISCO and gets ready for the next shrinkage. 
   It means more time to arrange for additional satellites/observatories aiming at observing the next, upcoming jet ejection.

In the case of 
XTE~J1859, 
for example, a half-day 
can be spent for discussing whether 
ToO observations are worthwhile to try. 
If the rapid, decreasing trend of both $R_{\rm in}$ and rms 
(as well as the spectral hardness) 
continues 
and 
if $R_{\rm in}$ almost reaches the ISCO, 
then an observation can be started. 
For the remaining three sources, one has 
more time to make 
a decision and prepare 
observations. 
A development of the ``jet forecast'' enables us concentrate 
observing facilities on the expected timing of jet ejections, and 
accelerates our understanding of jet physics.

The monitoring of $R_{\rm in}$
is more efficient 
than rms and the Hardness-Intensity diagram, 
in predicting when a next jet is launched. 
In the Hardness-Intensity diagram, 
there are many epochs near the ``jet line'' that do not 
show radio activities by the next exposure. 
In terms of 
rms, 
the threshold of rms to launch jets is uncertain. 
On the other hand, there is 
a reference point (the minimum value) 
for 
$R_{\rm in}$, 
which 
likely 
corresponds to 
the ISCO.

If the rapid rotation of BHs 
[high spin parameters $a$ or high absolute values $|a|$; 
such as, e.g., $\gtrsim 0.5$; \cite{Garofalo2009}] are 
essential to launch jets \citep{BZ1977}, 
some sources, harboring slowly-rotating BHs, 
will show little (or no) radio activity 
even when they meet the conditions of 
$R_{\rm in}$ (i.e., a rapid shrinkage of $R_{\rm in}$ 
down to the ISCO).
Thus, systematic investigations 
will 
answer whether the BH spin 
is 
the key for jet formations or not. 
If 
comprehensive studies for tens of BHs 
find 
no such sources, 
 a paradigm of the spin-based jet formation will 
face a challenge.

Since the accurate determination of $T_{\rm in}$ is crucial 
for 
precise measurements of $R_{\rm in}$, 
this ``jet forecast'' 
may be 
less effective for sources with 
relatively 
low-temperature disks, 
such as 
massive BHs.
The large effective area at 
low energies and 
wide-band coverage (to estimate the contribution 
from the hard component), 
as well as 
monitoring capabilities over numerous sources, 
are thus essential for this purpose.

Other than the conditions for jet launching discussed in 
this work (a shrinkage of $R_{\rm in}$), 
jets are occasionally formed after sudden decreases of the 
X-ray fluxes in 
GRS~1915$+$105 \citep{Mirabel1998} 
and 
3C120 \citep{Marscher2002}. 
Careful investigations are necessary to figure out 
whether the two different 
conditions are really working for jet launching.

\begin{acknowledgments}
We are grateful 
to the anonymous referee for helpful comments that greatly improved the paper,  
to Satoshi Nakahira for providing the MAXI data, and 
to Christian Knigge, Chris Done, and Mariko Kimura  
for helpful comments in the 
87th Fujihara Seminar 
``The 50th Anniversary Workshop of the Disk Instability Model 
in Compact Binary Stars''. 
We also appreciate 
Michael L.\ McCollough, 
Ruben Farinelli, 
Sergei Trushkin, 
 and 
Hitoshi Negoro 
for fruitful comments. 
TK acknowledges the support 
by the Thirty Meter Telescope (TMT) Project,
    National Astronomical Observatory of Japan, 
    through its funding program for research and development 
    of TMT science instruments,
by the joint research program 
of the Institute for Space Earth Environmental Research (ISEE), 
Nagoya University, 
    and 
by JSPS KAKENHI Grant Number 25K07370.
\end{acknowledgments}





%
\facilities{RXTE(PCA and HEXTE), NICER, MAXI(GSC), MOST, ATCA}

%
\software{
HEASoft \citep{Heasoft2014},
XSPEC \citep{Arnaud1996}
          }



\bibliographystyle{aasjournalv7}


\end{document}